\documentstyle[12pt]{article}
\makeatletter
\@addtoreset{equation}{section}
\makeatother

\begin{document}
\begin{center}
{\Large \bf
Indeterministic Quantum Gravity and Cosmology} \\[0.5cm]
{\large\bf XI. Quantum Measurement}\\[1.5cm]
{\bf Vladimir S.~MASHKEVICH}\footnote {E-mail:
mash@gluk.apc.org}  \\[1.4cm]
{\it Institute of Physics, National academy
of sciences of Ukraine \\
252028 Kiev, Ukraine} \\[1.4cm]
\vskip 1cm

{\large \bf Abstract}
\end{center}

This paper is a sequel to the series of papers [1-10]. We
define a quantum measurement as a sequence of binary quantum
jumps caused by a macroscopic apparatus. A dynamical theory
of measurement is developed, the role of gravity and
cosmology being emphasized.

\newpage

\hspace*{10 cm}
\begin{minipage} [b] {9 cm}
Measurement began our might.
\end{minipage}
\begin{flushright}
W. B. Yeats \vspace*{0.8 cm}
\end{flushright}

\begin{flushleft}
\hspace*{0.5 cm} {\Large \bf Introduction}
\end{flushleft}

Quantum theory has faced the problem of quantum measurement
for more than 70 years. There is voluminous literature
devoted to this problem, which cannot be reviewed here.
As for a critical part, we restrict our consideration to
three aspects of the problem of quantum measurements and
quantum jumps in general: energy conservation, decoherence,
and nonlocality. We treat a quantum measurement in the light
of quantum jump dynamics developed in this series of papers
[1-10].

The transformation of a pure state into a mixed one results
generally in changing the mean value of the Hamiltonian, so
that had a measurement of any observable in any state been
possible, the law of conservation of energy would have
been violated. Then it would have been possible to construct
perpetuum mobile of the third kind---a device increasing
energy and entropy. The impossibility of such a device imposes
essential restrictions on measurements and quantum jumps in
general.

The concept of decoherence claims that it is the latter that
causes the transformation of a pure state into a mixed one.
But as long as the state amplitude is a vector in a separable
Hilbert space, this claim is untenable. Therefore a theory of
measurement and quantum jumps in general must not be based on
the concept of decoherence.

Quantum entanglement makes possible a situation where a local
measurement results in a nonlocal quantum jump. Thus the issue
of the relationship between quantum jumps and relativity theory
arises. This issue cannot be resolved within the framework
of special relativity.

We define a quantum measurement as a sequence of binary
quantum jumps
caused by a macroscopic apparatus. As for the content of the
present paper, the main results are as follows.

In view of the principle of cosmic energy determinacy [1], the
law of conservation of energy holds at quantum jumps, so that
there is no perpetuum mobile.

Quantum jump dynamics [5] has nothing to do with decoherence, so
that the concept of the latter is denounced.

Nonlocality of quantum jumps implies an additional structure of
the spacetime manifold, which is absent in special relativity.
The structure is this: The hypersurface of a quantum jump is that
of a constant value of the cosmic time [1,2,5].

Any quantum jump is binary [5], so that a measurement is, in fact,
a sequence of binary jumps.

A macroscopic apparatus causes a quantum jump via the effect of
gravitational autolocalization [8,9], due to which the apparatus
pointer takes up a definite position.

\section{Measurement as quantum jumps}

We define a quantum measurement as a sequence of binary quantum
jumps caused by a macroscopic apparatus. One binary jump is an
elementary measurement.

Let $A$ be an observable measured in a system,
\begin{equation}
A=\sum_{i}a_{i}P_{i},
\label{1.1}
\end{equation}
where $P_{i}$'s are projectors. Let $\varphi,\;\psi$, and $\Psi$
be state vectors relating to the system under measurement, the
apparatus, and the composed system respectively. We have:
\begin{equation}
{\rm before\; the\; measurement}\qquad
\Psi_{b}=\varphi\otimes\psi,
\label{1.2}
\end{equation}
\begin{equation}
{\rm after\; the\; measurement}\qquad \Psi_{ai}=\varphi_{i}
\otimes\psi_{i}\quad {\rm with\; a\; probability}\quad w_{i}
\label{1.3}
\end{equation}
where
\begin{equation}
\varphi_{i}=\frac{P_{i}\varphi}{\|P_{i}\varphi\|},
\qquad w_{i}=(\varphi,P_{i}\varphi).
\label{1.4}
\end{equation}

Statistical operators of the composed system before and after
the measurement are
\begin{equation}
\rho_{b}^{\rm comp}=P^{\rm comp}_{\Psi_{b}}
\label{1.5}
\end{equation}
and
\begin{equation}
\rho^{\rm comp}_{a}=\sum_{i}w_{i}P^{\rm comp}_{\Psi_{ai}}
\label{1.6}
\end{equation}
respectively, where $P_{\Psi}^{\rm comp}$ is a projector
corresponding to a vector $\Psi$ of the composed system.

\section{Against perpetuum mobile of the third kind}

The mean value of the energy of the composed system is
\begin{equation}
E={\rm Tr}\{\rho^{\rm comp}H\},
\label{2.1}
\end{equation}
so that the change of the energy is
\begin{equation}
\Delta E\equiv E_{a}-E_{b}=\sum_{i}w_{i}
(\Psi_{ai},H\Psi_{ai})-(\Psi_{b},H\Psi_{b}).
\label{2.2}
\end{equation}
The change of entropy is
\begin{equation}
\Delta\sigma\equiv\sigma_{a}-\sigma_{b}=\sigma_{a}=
-\sum_{i}w_{i}\ln w_{i}>0.
\label{2.3}
\end{equation}
In the case of measuring an arbitrary observable $A$ we obtain
\begin{equation}
\Delta E\ne 0,
\label{2.4}
\end{equation}
i.e., the violation of the law of conservation of energy.

According to the first law of thermodynamics
\begin{equation}
\Delta E=Q+W
\label{2.5}
\end{equation}
where
\begin{equation}
Q={\rm Tr}\{(\Delta\rho^{\rm comp})H\}
\label{2.6}
\end{equation}
is heat and
\begin{equation}
W={\rm Tr}\{\rho^{\rm comp}(\Delta H)\}
\label{2.7}
\end{equation}
is work. In our case, eq.(\ref{2.2}),
\begin{equation}
\Delta E=Q.
\label{2.8}
\end{equation}
Let $Q>0$, so that
\begin{equation}
Q>0,\;\;\;\Delta\sigma>0.
\label{2.9}
\end{equation}
Such a self-heating system may be called perpetuum mobile
of the third kind.

The requirement that the equality
\begin{equation}
\Delta E=0
\label{2.10}
\end{equation}
must hold imposes essential restrictions on measurements
and quantum jumps in general.

In indeterministic quantum gravity and cosmology (IQGC), in
view of the principle of cosmic energy determinacy [1], the
law of conservation of energy holds at quantum jumps, so
that there exists no perpetuum mobile.

\section{Against decoherence concept}

By the concept of decoherence, the pure state
\begin{equation}
\tilde \rho_{a}^{\rm comp}=P^{\rm comp}_{\Psi_{a}},\qquad
\Psi_{a}=\sum_{i}(P_{i}\varphi)\otimes\psi_{i},
\label{3.1}
\end{equation}
of the composed system is equivalent to the state
$\rho_{a}^{\rm comp}$ given by eq.(\ref{1.6}),
\begin{equation}
\tilde\rho_{a}^{\rm comp}=\rho_{a}^{\rm comp},
\label{3.2}
\end{equation}
eq.(\ref{3.2}) being fulfilled due to the decoherence condition:
\begin{equation}
(\Psi_{ai'},O\Psi_{ai})=0 \quad {\rm for}\quad i'\ne i
\label{3.3}
\end{equation}
where $\Psi_{ai}$ is given by eq.(\ref{1.3}) and $O$ is any
observable of the composed system.

In IQGC, quantum jump dynamics [5] has nothing to do with the
absurd condition given by eq.(\ref{3.3}), so that the concept
of decoherence is denounced.

\section{Against overlooking gravity and cosmology in quantum
jump dynamics}

To conceive of the role of gravity and cosmology in quantum
jump dynamics, let us consider a geometric, or coordinate-free
description of quantum fields. In the Heisenberg picture, a
quantum field is
\begin{equation}
\phi_{H}\equiv\phi=\phi(p), \qquad p\in M,
\label{4.1}
\end{equation}
where $M$ is a spacetime,
\begin{equation}
\Psi_{H}\equiv\Psi
\label{4.2}
\end{equation}
is a field state vector;
\begin{equation}
\phi^{class}(p)=(\Psi,\phi(p)\Psi)
\label{4.3}
\end{equation}
is a classical field.

In the Heisenberg picture, the state vector $\Psi$ changes at
and only at quantum jumps. To every quantum jump there corresponds
a spacelike hypersurface. The hypersurfaces must be mutually
disjoint: otherwise $\Psi$ would be not defined. The quantum-jump
hypersurfaces are causewise ordered:\begin{equation}
{\cal S}_{2}>{\cal S}_{1},\;\;{\rm or}\;\;{\cal S}_{1}<{\cal S}_{2}.
\label{4.4}
\end{equation}
Within the limits of special relativity, i.e., if $M$ is the
Minkowski spacetime, it is neither theoretically nor
experimentally possible to determine what are those hypersurfaces
and which is, in the general case, their causelike order. It is
special relativity that prevents a complete phenomenological
mathematical description of quantum jumps and, by the same token,
measurements.

A complete dynamical description of quantum jumps has been given
in IQGC. The quantum jumps imply an additional structure of the
spacetime manifold, which is absent in special relativity. The
hypersurface of a quantum jump is that of a constant value of the
cosmic time $t$, so that
\begin{equation}
{\cal S}_{2}>{\cal S}_{1}\Leftrightarrow t_{2}>t_{1}.
\label{4.5}
\end{equation}

\section{Binary jumps and measurement}

By eqs.(\ref{1.1}),(\ref{1.4}) we have for a measurement:
\begin{equation}
\{P_{i}:i\in {\cal I}\},\;\;\;\sum_{i\in {\cal I}}P_{i}=I,
\label{5.1}
\end{equation}
\begin{equation}
w_{i}={\rm Tr}\{\rho P_{i}\},\;\;\;\rho\equiv\rho_{b}=P_{\varphi}.
\label{5.2}
\end{equation}
An elementary measurement is described by a binary quantum jump,
for which the equation
\begin{equation}
P^{+}+P^{-}=I
\label{5.3}
\end{equation}
holds [5]. Now an equation
\begin{equation}
P_{i}=P^{1s_{1}\:2s_{2}\ldots ns_{n}},\;\;\;n=n(i),
\;\;\;s_{k}=\pm,
\label{5.4}
\end{equation}
and relations
\begin{equation}
P^{1s_{1}\ldots (n-1)s_{n-1}\:ns_{n}}P^{1s_{1}\ldots
(n-1)s_{n-1}}=P^{1s_{1}\ldots ns_{n}},
\label{5.5}
\end{equation}
\begin{equation}
P^{1s_{1}\ldots (n-1)s_{n-1}\:n+}+P^{1s_{1}\ldots
(n-1)s_{n-1}\:n-}=P^{1s_{1}\ldots (n-1)s_{n-1}}
\label{5.6}
\end{equation}
hold.

We have for the probability of the $i$-th result as the
sequence given by eq.(\ref{5.4}):
\begin{equation}
\tilde w_{i}={\rm Tr}\{\rho P^{1s_{1}}\}
{\rm Tr}\{\rho^{1s_{1}}P^{1s_{1}\:2s_{2}}\}\cdots
{\rm Tr}\{\rho^{(n-1)s_{n-1}}P^{1s_{1}\ldots ns_{n}}\}
\label{5.7}
\end{equation}
where
\begin{equation}
\rho^{ks_{k}}=\frac{P^{1s_{1}\ldots ks_{k}}
\rho^{(k-1)s_{k-1}}P^{1s_{1}\ldots ks_{k}}}
{{\rm Tr}\{\rho^{(k-1)s_{k-1}}P^{1s_{1}\ldots ks_{k}}\}}.
\label{5.8}
\end{equation}
We find
\begin{equation}
\tilde w_{i}={\rm Tr}\{P^{1s_{1}\ldots ns_{n}}
\cdots P^{1s_{1}}\rho P^{1s_{1}}\cdots P^{1s_{1}
\ldots ns_{n}}\}=
{\rm Tr}\{\rho P^{1s_{1}\ldots ns_{n}}\}=
{\rm Tr}\{\rho P_{i}\}=w_{i}.
\label{5.9}
\end{equation}

\section{Determinacy of the pointer position}

We take a ball as a model of the pointer of a macroscopic
apparatus and use the results [8,9] for the ball.

In IQGC the following relations take place:
\begin{equation}
H=H[g,\dot g]
\label{6.1}
\end{equation}
and
\begin{equation}
\ddot g=\ddot g[g,\dot g,\Psi]
\label{6.2}
\end{equation}
where $g$ is the metric and dot stands for the derivative with
respect to the cosmic time $t$. In the Newtonian approximation,
on the other hand,
\begin{equation}
H=H[\Psi],
\label{6.3}
\end{equation}
so that the approximation is inapplicable at the jump time
$t_{{\rm jump}}$.

Let at $t=t_{{\rm jump}}-\tau_{-}$ the relations
\begin{equation}
H\Psi=\epsilon\Psi,\qquad H\Psi'=\epsilon'\Psi'
\label{6.4}
\end{equation}
hold where
\begin{equation}
\Psi=(c^{+}\varphi^{+}+c^{-}\varphi^{-})\otimes\psi^{0},
\qquad \Psi'=(c^{-*}\varphi^{+}-c^{+*}\varphi^{-})
\otimes\psi^{0},
\label{6.5}
\end{equation}
\begin{equation}
\psi^{0}=\psi^{0}(\vec r),
\label{6.6}
\end{equation}
$\Psi$ is an actual state vector, and the quantum jump
is caused by the crossing of levels $\epsilon$ and $\epsilon'$.

Let at $t=t_{{\rm jump}}+\tau_{+}$ the relations
\begin{equation}
H^{\pm}\Psi^{\pm}=\epsilon^{\pm}\Psi^{\pm},\qquad
H^{\pm}=H^{\pm}[\Psi^{\pm}],
\label{6.7}
\end{equation}
hold, where
\begin{equation}
\Psi^{\pm}=\varphi^{\pm}\otimes \psi^{\pm},
\label{6.8}
\end{equation}
\begin{equation}
\psi^{\pm}(\vec r)=\psi^{0}(\vec r-\vec R^{\pm}),
\label{6.9}
\end{equation}
and
\begin{equation}
R=|\vec R^{+}-\vec R^{-}|\ll a_{0}.
\label{6.10}
\end{equation}
Consider $H[\tilde\Psi]$ where
\begin{equation}
\tilde\Psi=c^{+}\Psi^{+}+c^{-}\Psi^{-}.
\label{6.11}
\end{equation}
In view of eq.(\ref{6.10}), $V(\vec r;\tilde\Psi)$ has only
one minimum, so that $\tilde\Psi$ is not an eigenvector of
$H[\tilde\Psi]$. Thus the quantum jump takes place
with the result
\begin{equation}
\Psi^{+}\quad {\rm or}\quad \Psi^{-},\qquad w^{\pm}=
|c^{\pm}|^{2}.
\label{6.12}
\end{equation}
To the vectors $\psi^{\pm}$ there correspond two different
positions $\vec R^{\pm}$ of the pointer. Here no concept of
decoherence is involved.

\section*{Acknowledgment}

I would like to thank Stefan V. Mashkevich for helpful
discussions.


\begin{thebibliography}{99}

\bibitem{1} Vladimir S. Mashkevich, {\it Indeterministic
Quantum Gravity} (gr-qc/9409010, 1994).
\bibitem{2} Vladimir S. Mashkevich, {\it Indeterministic
Quantum Gravity II. Refinements and Developments}
(gr-qc/9505034, 1995).
\bibitem{3} Vladimir S. Mashkevich, {\it Indeterministic
Quantum Gravity III. Gravidynamics versus Geometrodynamics:
Revision of the Einstein Equation} (gr-qc/9603022, 1996).
\bibitem{4} Vladimir S. Mashkevich, {\it Indeterministic
Quantum Gravity IV. The Cosmic-length Universe and the
Problem of the Missing Dark Matter} (gr-qc/9609035, 1996).
\bibitem{5} Vladimir S. Mashkevich, {\it Indeterministic
Quantum Gravity V. Dynamics and Arrow of Time}
(gr-qc/9609046, 1996).
\bibitem{6} Vladimir S. Mashkevich, {\it Indeterministic
Quantum Gravity and Cosmology VI. Predynamical Geometry
of Spacetime Manifold, Supplementary Conditions for Metric,
and CPT} (gr-qc/9704033, 1997).
\bibitem{7} Vladimir S. Mashkevich, {\it Indeterministic
Quantum Gravity and Cosmology VII. Dynamical Passage through
Singularities: Black Hole and Naked Singularity, Big Crunch
and Big Bang} (gr-qc/9704038, 1997).
\bibitem{8} Vladimir S. Mashkevich, {\it Indeterministic
Quantum Gravity and Cosmology VIII. Gravilon: Gravitational
Autolocalization} (gr-qc/9708014, 1997).
\bibitem{9} Vladimir S. Mashkevich, {\it Indeterministic
Quantum Gravity and Cosmology IX. Nonreality of Many-Place
Gravitational Autolocalization:
Why a Ball Is Not Located in Different Places
at Once} (gr-qc/9802016, 1998).
\bibitem{10} Vladimir S. Mashkevich, {\it Indeterministic
Quantum Gravity and Cosmology X. Probability-Theoretic
Aspect: A Hidden Selector for Quantum Jumps, or How the
Universe Plays the Game of Chance} (gr-qc/9802022, 1998).
\end{thebibliography}
\end{document}